# Assessing the Value of Peer-Produced Information for Exploratory Search

Elizeu Santos-Neto[*], Flavio Figueiredo[+], Nigini Oliveira[#]
Nazareno Andrade[#], Jussara Almeida[+], Matei Ripeanu[*]

| University of British Columbia | Universidade Federal de Minas Gerais[+] | Univ. Federal de Campina Grande[#] |
|---|---|---|
| Elec. & Computer Eng. Department | Dep. de Ciência da Computação | Dep. de Sistemas e Computação |
| {elizeus,matei}@ece.ubc.ca | {flaviovdf,jussara}@dcc.ufmg.br | {nigini,nazareno}@lsd.ufcg.edu.br |

**ABSTRACT**
Tagging is a popular feature that supports several collaborative tasks, including search, as tags produced by one user can help others finding relevant content. However, task performance depends on the existence of 'good' tags. A first step towards creating incentives for users to produce 'good' tags requires the quantification of their value in the first place. This work fills this gap by combining qualitative and quantitative research methods. In particular, using contextual interviews, we first determine aspects that influence users' perception of tags' value for exploratory search. Next, we formalize some of the identified aspects and propose an information-theoretical method with provable properties that quantifies the two most important aspects (according to the qualitative analysis) that influence the perception of tag value: the ability of a tag to reduce the search space while retrieving relevant items to the user. The evaluation on real data shows that our method is accurate: tags that users consider more important have higher value than tags users have not expressed interest..

**Author Keywords**
Peer production; information value; tagging; exploratory search.

## 1. INTRODUCTION

Social tagging is a rich source of user-generated content. Although tagging started as a central feature of social bookmarking and media sharing tools such as *delicious.com*, *Flickr*, and *YouTube*, currently most major online social systems, such as *Twitter*, *Facebook*, and *Google+*, have adopted their own flavour of tagging as a part of their palette of features designed to engage users. These tags (or hashtags) enable users to annotate the content they (or others) have published in the system and to navigate the existing content. This continued interest in social tagging highlights the need to understand users' perception of tag value and to design methods that can quantify tags' value so that it can be used as a signal to other mechanisms. For instance, quantifying the value of tags from the perspective of an information seeker can enable the creation of incentive mechanisms to improve the overall quality of tags users produce. Additionally, tags can be used to improve the efficiency of many applications such as search [36], taxonomy generation [17], public library cataloguing [28], and clustering/classification [25].

In this setting, this work focuses on the problem of estimating the value of tags as perceived by users. This is key to incentivizing contributions and thus designing a sustainable ecosystem in the context of (online) *peer-production systems*: systems whose mode of producing goods is decentralized, non-proprietary, and collaborative [6]. At the heart of any such system is the notion that users produce value to their peers via individual contributions. Although social tagging systems are instances of peer-production systems, quantifying the *value* of individual contributions is an issue neglected by previous studies.

The problem of quantifying user contributions in peer-production systems where users share physical resources (e.g., car seats in carpooling, bandwidth in BitTorrent) is relatively simple as it largely reduces to counting the resource units one user donates. In information-sharing systems, however, the value of contributions is harder to assess, as the value of a piece of information is contextual and multidimensional [18,29]. More specifically, in social tagging systems, the value of a tag can vary widely from user to user, and even to the same user across applications/contexts (e.g., some tags may be good to categorize items, while others are better to discover new items) [32].

To make progress towards designing incentives for users to produce higher value contributions, this work focuses on the context of exploratory search (e.g., navigation based on tag clouds [13,23,31]), a common information retrieval task performed in social tagging systems. In this context, this study tackles two problems: first, *it characterizes the information seekers' perception of tags' value;* and second, *it designs a method to quantify the value of tags from the perspective of information seekers based on the aspects that influence their perception of value*.

*Context*. Exploratory search is a more general concept that encompasses tag-based navigation [21,23,34]. In its tag-based realization, exploratory search process can be described as follows, *information seekers* navigate the set of items by using tag clouds, as opposed to traditional keyword search. Users prefer tag-based navigation when they are exploring a topic and want to retrieve a set of related items, as opposed to the single most relevant item [27]. Tag clouds are the default interaction mode provided by systems like

*delicious.com*, *StackOverflow*, or *MrTaggy* [19]. Tag clouds are produced using the tags previously entered in the system by *information producers*. *Information seekers* start the navigation by entering a tag-query (typing or clicking). The system, in turn, retrieves items that are annotated with that tag-query and related tags (e.g., in the form of a tag cloud). The navigation continues further if the user selects one of the available tags presented by the system. The search result at each navigation step is composed of items annotated by all the tags selected by the user. Thus, we assume that the search provides AND-semantics [5].

***Research questions***. This work focuses on the following research questions:

*Q1. What are the aspects that influence users' perception of a tag's value for exploratory search?*

*Q2. How to quantify the value of tags as perceived by information seekers in exploratory search?*

To address both questions, we use a combination of qualitative and quantitative research methods. In particular, we use contextual interviews and grounded theory methods [15] to characterize the aspects that influence users' perception of tag value (Q1); and, define metrics that capture such aspects, using analytical proofs and experiments with tagging data to evaluate the defined metrics (Q2).

In summary, *this work makes three major contributions*:

*i)* A qualitative characterization of users' perception of tag value in the context of exploratory search, based on contextual interviews. This reveals that the two salient aspects that influence users' perception of tag value are: *ability to retrieve relevant content items* and *ability to reduce the search space*. Additionally, it presents an investigation of other aspects of tag production that contribute to solidify the existing body of research on users' motivation behind tagging. (Section 3)

*ii)* A method to automate tag value estimation that caters for the two desirable tag properties in the context of exploratory search, as identified by the qualitative user study. We prove that this method has desirable theoretical properties while quantifying these two aspects. (Section 4)

*iii)* A validation experiment using real tagging datasets that shows that the proposed method accurately quantifies the value of tags according to users' perception (Section 5).

## 2. BACKGROUND AND RELATED WORK

In a nutshell, this work differs from previous efforts in two aspects: first, it is motivated by the view that social tagging systems are inherently online peer production systems. Thus to improve the quality of user contributions, it is necessary to first quantify their value, so that one can then think of designing incentives for the production of high quality content. Second, this work focuses both on characterizing users' *perception* of tags' value, and on the design/analysis of a method to assess tag value in practice.

### 2.1. Why Do We Tag?

Hammond et al. [11] provide, perhaps, the first study that discusses the characteristics of social tagging, its potential, and the motivations users have to produce tags. The study comments on the features provided by different social tagging systems, and discusses preliminary reasons that incentivize users to annotate and share content online.

Marlow et al. [24] discuss the properties of several tagging systems while pointing out their similarities and differences. Additionally, the authors conjecture the motivations that can potentially drive the production of tags. Ames and Naaman [2] go deeper on the study of motivations behind tagging and investigate why people tag in mobile (i.e., *ZoneTag*) and web applications (i.e., *Flickr*). They interviewed 13 users to address the question: 'Why do people tag?' Their findings indicate that there are both personal and social motivations behind tagging.

*Our work* differs from these previous studies as it concentrates on understanding the use (i.e., consumption) of tags to engage in exploratory search tasks (e.g., exploring the set of items available in a social bookmarking tool), as opposed to focusing on the motivations behind the production of tags.

### 2.2. Contributions in Peer Production Systems

Online peer production systems can be categorized into systems where users produce/share *resources* or *information*. In the former category, as we have already mentioned, quantifying the value of user contribution is based largely on counting the resource units one user produces and donates to other users (and implicitly to the system). For example, in P2P content sharing systems (e.g., BitTorrent) the value of contributions is estimated by the volume of content a peer donates to others [3].

Valuing contributions in these resource-sharing peer production systems relies on: first, the fact that the amount of resources donated are easily quantifiable (e.g., CPU count or bandwidth); second, the assumption that contribution value can be directly linked to the resources consumed to deliver a service; and, third, on the simplifying assumption that a unit of contributed resources has a uniformly perceived value across all users.

In contrast, none of these assumptions hold for systems that support production/sharing of information. First, it is impossible to directly quantify the 'effort' that has led to the production of a specific piece of information; and, second, the value of information (e.g., tags or items in tagging systems) is subjective to users' interests, and task at hand (an aspect shared with other information goods).

*In this study*, we cope with the contextual nature of tag value by: first, using a qualitative analysis to identify the aspects that influence information seekers' perception of tag value (in the context of exploratory search); and, second, using the result of this analysis to inform the design of a method that quantifies the value of tags.

### 2.3. Characterizing the Quality of Tags

Several studies focus on characterizing the *quality* of tags or tagging (as a feature of an information system) in general. These studies instantiate the notion of 'quality' in various ways, which we comment in turn.

*Search and Recommendation*. Focusing on the quality of tags for information retrieval tasks such as content classification and search Figueiredo et al. [9] and Bischoff et al. [7] evaluate the quality of information provided by tags (in comparison to other textual features) to improve the efficiency of recommendation mechanisms. Similar studies aim to harness tags to improve web search [16,36]. In the context of decentralized search, Helic et al. [13] studies tagging systems from a network theoretical perspective by analyzing whether tagging networks (i.e., formed by connecting tags to items) have properties that enable efficient decentralized search [1]. In particular, the authors study the impact of different methods that build such tag networks (i.e., tag hierarchies, *folksonomies*) on a decentralized search process.

*Tagging as a categorization mechanism*. In a different application, Heymann et al. [17] investigate whether tags help users to categorize content by analogy with widely deployed classification tools for library management. They use a qualitative analysis to evaluate the power of tags to build classification systems rather than a user-centric quantitative approach to assess value. Lu et al. [22] perform a similar study, by comparing peer-produced tags and expert-assigned terms to classify books, showing that tags can improve accessibility of items in a library catalog.

*Quality of textual content*. Other studies focus on the contents carried by tags. Suchanek et al. [30] study the quality of tags by determining the descriptive power of a tag (i.e., its efficiency in describing an item). Similarly, Gu et al. [10] propose a way to measure the *confidence* in which a tag describes a particular item. In their context, confidence equates to the relevance of the tag to the topic of item. More recently, Beza-Yates et al. [4] characterize the lexical quality of several web sources, including a social tagging system (Flickr.com), finding that the lexical quality of texts in Flickr is better than that in the general web. Other work has focused on methods to detect and mitigate the impact of tag spam [20].

*Building tag clouds*. Helic et al. [14] and Venetis et al. [31] analyze algorithms to build tag clouds. Their approach to evaluate the quality of a tag cloud is directly related to ours: such studies resort to metrics that aim to capture intuitive aspects of users' information needs (novelty, diversity, and coverage). Complementary to the study of tag clouds quality, Wilson et al. [35] investigate whether tag clouds are valuable for sense making, with their findings suggesting that tag clouds have little to no effect on the process of sense making.

*Our approach* differs from previous efforts as we start by characterizing users' perception of tag value to inform the design of a method that quantifies the value of tags for exploratory search. To the best of our knowledge there are no previous attempts to neither characterize the perception of tag value for exploratory search nor design methods to quantify tag value in such context.

### 2.4. Economics of Information

Hischleifer [18], while studying the characteristics of information goods, enumerates and discusses a set of economically significant information attributes that can influence its perceived value, namely: *Certainty*, *Diffusion, Applicability, Content,* and *Decision-relevance*. Moreover, the value of information in market settings is contextual [29], as it requires one to make use of it to assess its expected value. Repo [26] goes further with this statement and discusses two major approaches on assessing value of information: value-in-use and exchange value. While the attributes that influence information value have been investigated in market contexts, it is unclear what role these attributes play, if any, in the context of peer-production systems, and in particular on social tagging systems.

*Our study* uses the same notion of a multidimensional value concept as discussed by Hischleifer. We inquire further on the human perception by performing interviews with users to understand their perception of value of peer-produced information (*which departs from the type of information focused on in previous work*). More precisely, we investigate what aspects users take into account when choosing tags in the context of exploratory search tasks. Understanding the value of information in online peer production systems, as perceived by information seekers, extends the existing body of knowledge on the value of information in markets and in the design of information systems discussed in the next section.

### 3. USERS' PERCEPTION OF VALUE – A QUALITATIVE INVESTIGATION

This section presents a *qualitative* investigation on the aspects that influence users' perception of tag value in social tagging systems, focusing on exploratory search tasks. To contribute to the existing body of research on the motivation behind tagging, we also present the findings about aspects related to the users' *production* of tags.

*Participants*. The target population for this experiment was Internet users who are familiar with search and navigation tasks in social tagging systems. With this mindset, subjects were recruited through a combination of advertisement via email and snowball recruiting techniques which led to 12 participants. Two initial interviews were used as pilots to refine the interview protocol and were discarded from the final analysis. One other interview was discarded as the she failed to display basic familiarity with social tagging.

Analyzing the remaining samples suggests that saturation in the information obtained from the interviews was reached.

Participants were asked to complete a background and demographics questionnaire, and are mostly young males: only one is female and only two reported to be over 30 years old (two did not report their age). Brazilians (5), Iranians (2) and USA nationals (2) compose the group which is highly educated: all of them have at least a post-secondary degree. The majority of the participants has an engineering/computer science background, while two others have background in linguistics and arts. All participants reported to be fully capable of performing exploratory search tasks, and eight reported to be able to develop software.

*Data collection and analysis*. The data was collected using semi-structured contextual interviews, which provides flexibility in approaching participants about their tag-based search habits. The interview protocol consisted of open-ended questions that explore the users' application of tagging features in different systems (See Appendix C).

Each interview lasted for one hour, and consisted of two parts. Both parts consisted of contextual inquiries where participants were encouraged to use a social tagging system to illustrate usage and explain their choices of tags while searching. In the first part of the interview, participants were free to use any system they were familiar with. The goal was to gain an insight into their habits as they explained their understanding of tags and their personal usage choices.

In the second part, the participant used a *Delicious*-clone system we have created. This is a bookmark system that is searchable by directly typing or clicking on tags (which are presented on a tag cloud and as related tags to returned items), where results are presented in a paginated list. The system was populated with a snapshot of bookmarks and tags from *Delicious* containing more than 600K entries collected in September 2009. During and after his usage, the interviewer posed questions. The goal of this task-driven interview is to inquire deeper on the users` decision making when performing exploratory searches while being able to record users' interaction with the system (i.e., collect click-traces). By motivating the user with a real search task that is similar to the user` common tasks, we can explore specific aspects that influence the choice for one tag versus another.

All interviews were performed face-to-face or using a video chat tool. All sessions were recorded as a video of the participant's screen and the audio of the conversation. The data collected via the interviews were transcribed verbatim, coded, and, finally, analyzed using Grounded Theory methodology [15]. Initially a pair of researchers separately coded two interviews each and together used this data to build a codebook containing both inductive and deductive codes. All interviews were coded and the codebook reviewed when necessary. Memoing was used all time to keep track of issues that were discussed at most in weekly meetings. An analysis plan was used to aggregate the discussion data and its analysis based on interview excerpts. Finally thick descriptions of each topic discussed in the next sections were built considering emerging patterns through code categorization.

### 3.1. Aspects of Tag Production
While our main goal is to characterize the perception of tag value in exploratory search tasks to inform the design of methods that quantify tag value, we also probe what aspects influence users' perception of tag value when producing annotations (as opposed to using tags to search).

A prevalent theme observed in the interviews is that users perceive tags as valuable when they help describing items they are annotating, and thus improve sense-making about a set of items and by making individual items searchable. In particular, interviewees comment on the need for tags to describe images and videos with these two purposes. In this context, *tags that describe features of the object* such as location, people, and aesthetics characteristics are considered useful (e.g., P4 told that she tagged a video with its soundtrack musician name because: "*imagine that someone is searching about (artist name)… they will find me and it will be great!*"). For *tweets*, which themselves are searchable, tags are reported as useful to augment their meaning by making explicit a *feeling* about the text or providing *context* for the textual item (e.g., P6 cites this tweet: '*20 minutes in a queue!* #angry' and explains that "*it was not to classify anything … I'm using a tag to express a feeling*").

While creating annotations to improve the ability to find the item later, some participants report that *there is a tension between using general and specific terms*. General tags are likely memorable but provide little discriminative power. P3 gives an example: "*If I didn't use an obvious tag, I'd not remember that* [an article that would be helpful]. *But this probably made me use too broad tags. There's probably some tags like 'programming'.*"

Another aspect repeatedly raised by our subjects is the potential of tags to attract attention to items they create or post, so that they would likely become more popular or more likely to be found. Participant P11 described a strategy to promote content by the use of tags: "*instead of writing 'got first place in the fencing championship', I write 'got first place in the #fencing champion' as it makes easier to others find my tweet when searching for that tag*".

Finally, interviewees commented how annotations may attach the content to a trend, or the contributor to a group. Annotating an item with a tag that is currently used in a trending topic or which is specific to groups is seen as connecting the user with others. For example, P4 on Instagram usage: "*(a famous user) creates this hashtags so everybody can submit stuff*"; and P11 on tweets about the Brazilian 2013 street protests: "*I'd avoid to use a tag that I know to be used by people with a different political position than mine*".

*In summary, the aspects that influence one user's perception of value during tags production may not be in tandem with the expectation of another user when searching for items*, as some of the driving forces behind the perception of value during tag production are highly personal (e.g., feelings). Thus, other users may not consider these tags valuable when trying to locate an item.

## 3.2. Users' Perceptions of Tag Value

Exploratory search is the process of acquiring a set of information resources that respond to an information need (e.g., a particular domain) with a certain level of certainty. This section presents the qualitative analysis of aspects that influence users' decision-making along the steps of exploratory search.

Users provided data about their decision-making either voluntarily or by answering specific questions about their actions while trying to locate items that fulfill their needs. Based on our observations the exploratory search process can be described as follows:

The user enters a loop (*i*) deciding which tags to use to *define a search space* at each step that most accurately reflects her information need; and (*ii*) judging the *relevance* of returned items; finally, the user leaves the loop by (*iii*) selecting items that satisfy their needs.

*Search space definition*. A search space is a set of items from which users can select a subset (i.e., define a subspace) via tags that annotate them. Participants normally define a search space by expressing their information needs via tags. Search space definition is an essential part of the exploratory search process and involves different perspectives of the set of items retrieved by each tag. Participant P11, for example, during the execution of a **Search Task 1** (see Appendix C[1]), clicked on '*web2.0*' and reported that this tag "*is more representative of web social networks*" (which was the main topic of that participant's information need). The same idea was expressed by participant P6 when choosing the tag '*tutorial*', which according to the participant, was a better representation of her particular information need (i.e., programming) than the other tags available at that exploration stage.

*Known vocabulary*. As users try to translate their information need into tags, these tags tend to come from users' *known vocabulary*. Participant P8 is clear about that when saying that she chooses "*hashtags that are alike terms that I hear*", when performing exploratory search on *Twitter*. The same user goes further and comments on the 'cryptic' aspect of the tag #DAADC13 saying: "*this one here I would probably not click on because I do not know what it means*". However, this is not simply a matter of a tag to be 'known or unknown' to a given user. Participant P3 justifies choosing 'computer_science' to search instead of 'computing' saying that "*basically it's because I use it more often*". These observations suggest that the more a tag is used by a user, the higher its perceived value is.

*Search space size*. Users tend to refer to the 'right size' of a search space (i.e., the number of items it contains) in exploratory search when talking about the decision to continue searching for items. Participant P11, for instance, mentions that "*A lot of results is confusing and you'll not be able to find what you want… a number of results that doesn't even fill system's first page is kind of frustrating*". Additionally, P4 expressed a "*lack of confidence*" in the results when "*too many items were retrieved*". In contrast, participant P1 took the action of removing an added tag because "*it might have filtered too much*". Interestingly, many participants mentioned the number of retrieved items (reported by the system in the search results page) as a way to gauge whether the tags are helping on controlling the search space size. As mentioned by the users, search space size affects their perception of the value of a tag.

*Relevance*. Besides finding the '*right size*' of a search space, no search is complete without locating relevant items. In fact, the *relevance* aspect has been raised and described by *all participants*, which strongly suggests that it is a major influence on users' value perception of value.

Participant P7 points out to this aspect by stating: "*I am going to take a look at the first five or ten entries to have an idea about my results*". After a brief inspection, the participant decides that "*they* (the results) *still have a lot of noise, so I am going to add one more tag*". Similarly, participant P8 reports an analysis of the relevance of the space defined by a tag as saying that "*this* (set of items) *is still not sufficient … I gave a quick look but the first* (entries) *were not interesting*". Participant P7 is more direct in suggesting that *relevance* influences the perceived value of a tag when reasoning about a particular choice of tags. The participant selected '*software*' instead of '*programming*' based on the perception that she "*will find more things related* (to my information needs)" using the former instead of the latter.

*Combination of space size and relevance of items*. Participants use words like '*focused*', '*specific*', '*restrict*', and '*refined*' to describe a desired search space that balanced well size and relevance. Participant P7 supports this observation by explaining a click decision: "*as 'opensource' is already a subset of* (software) *development/programming then I'll start clicking at 'opensource'*". Similarly, participant P1 reasons that adding an additional tag to the navigation is beneficial because "*it might give more focused results*". Another strong example related to the influence of the combinations of these two space characteristics – size and relevance of items – is raised by P3 when deciding to redefine the space at a particular point of the navigation: "*It looks like this (result) is really related to 'storage' but there is nothing to do with research. I need to refine it more*". This combination of characteristics of a space (as defined by a tag) influences positively a tag's value, as a tag can define both a *smaller space* that contains highly *relevant items*.

*Diversity and neighbouring spaces*. Finally, two other identified aspects are connected to tags related to a currently defined search space: *diversity* and *neighboring spaces*. To some degree, these two aspects are opposite concepts if one considers that related tags to a given space (presented as a tag cloud) can be perceived as increasing the diversity of items in that space or simply retrieving similar neighboring spaces.

Participant P4 considers confusing to have '*artists*' as part of the tag cloud when the current space is already defined by the tag '*artist*', which suggests that more diversity in the tag cloud improves the perception of value for the tags in the tag cloud relative to the currently defined space. Similarly, participant P1 is even more emphatic about this aspect while performing **Search Task 1** (see Appendix C) by stating that: "*type and typography both of them point to the same thing, web and website, icon and icons, it's a bit of useless to have these two similar, very similar tags together, this is something that impacts the value, icons have zero value here because you have icon here*". When inquired about whether replacing these highly similar tags by more diverse set of tags would improve the perceived value, the participant replied: "*Yes, meaningful diversity within the tags*".

On the other hand, participant P2 selects the tag '*user experience*' after using '*ux*', while reporting that these two terms are considered synonyms. Participant P2 explains that she perceives that the tag '*user experience*' can retrieve results similar to those retrieved (but not annotated with) by the tag '*ux*' (i.e., a neighboring space to the currently defined one). We were unable, however, to identify whether one of these two aspects is more important than the other regarding the characteristics of a tag cloud to users.

### 3.3. Discussion and Summary
The analysis reported in the previous section leads to several insights into the aspects that influence the users' perception of tag value. These insights are summarized by the concept map in Figure 1.

In particular, two aspects are more salient, as expressed by the participants: *search space size* and *relevance*. Therefore, our findings suggest that *the perceived value of a tag is largely influenced by its ability to retrieve items that are relevant to a user while reducing the search space size*. The tag reduces the search space by *filtering out items,* and maximizes relevance by *retaining the items that address the user's information needs* in exploratory search.

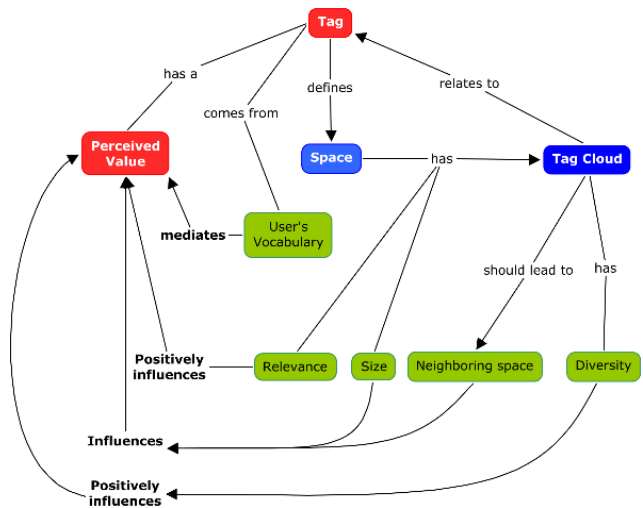

**Figure 1. Summary of aspects that influence the users' perception of tag value, as reported by the participants of our qualitative study.**

Finally, it is worth highlighting that this study provides an important characterization that can help designing new social tagging features (e.g., tag cloud algorithms, user interface design, and ranking mechanisms), as it improves our understanding of what users consider valuable when searching with tags.

## 4. QUANTIFYING TAG VALUE
This section introduces our method to estimate the value of tags. This method focuses on the two most salient aspects highlighted in previous section: *relevance* and *space size*.

Given a tag and an information seeker, there are multiple ways to formalize and quantify these two aspects. In this section, we propose a method and show analytically that it has desirable properties. In the next section we validate this approach by evaluating its impact it on real datasets.

### 4.1. System Model
Let $\mathbb{S} = (U, I, A)$ be a social tagging system, where $U$ represents the set of users in the system, $I$ denotes the set of items, and $A$ represents the set of annotations. An annotation is a tuple that specifies its author, the annotated item, a tag assigned to the item, and the time the annotation happened. Formally, $A = \{(s, i, t, e) | s \in U, i \in I, t \in T\}$, where $t$ is a *tag*, a word selected by the user from any vocabulary to annotate the item at timestamp $e$).

The set of annotations $A_s$ characterizes a particular user $s$, where individual annotations can be distinguished by their timestamps. More formally, $A_s = \{(q, i, t, e) \in A | q = s\}$. From the set of annotations $A_s$, it is possible to derive the set of items (or *user library*) $I_s$, and the set of tags (or *user vocabulary*) $T_s$, respectively annotated and used by particular user $s$, as follows: $I_s = \{i | (s, i, t, e) \in A_s\}$, and $T_s = \{t | (s, i, t, e) \in A_s\}$. The set of tags assigned to a particular item $i$, $T^i$, and the set of items tagged with a particular tag $t$, $I^t$, are similarly defined.

**Item relevance**. We assume that, for an information seeker $s$, there is a probability mass function $p(i|s)$ over the set of items in the system that the information seeker has not annotated yet (i.e., $I - I_s$) that specifies the relevance of an item $i$ given that information seeker $s$. Therefore, the set of items relevant to an information seeker $s$ can be defined as:

DEFINITION 1. *Given an information seeker $s$, the set of items relevant to $s$ is:* $\Gamma_s = \{i \in I - I_s | p(i|s) > 0\}$, *where $p(i|s)$ the probability of relevance of an item $i$ to an information seeker $s$.* □

Note that $\Gamma_s$ can be defined for different search tasks, such as exploring a user's own library (i.e., items already tagged by the user). The proposed method is general enough to work with alternative definitions of $\Gamma_s$.

*Modeling exploratory search*. We model exploratory search as a communication channel between the search engine (the sender) and an information seeker (the receiver). Consider that the sender transmits items to the user, and the channel is characterized by the probability of item relevance $p(i|s)$ to the receiving user $s$ over the set of items $\Gamma_s$. In this context, a tag defines a filter that creates a new channel from the original one: this new channel is characterized by the probability of item relevance conditional on that tag and on the seeker: $p(i|t,s)$. Thus, we model exploratory search in a probabilistic manner, as opposed to a deterministic one (e.g., simply counting the number of relevant items retrieved by each tag).

*Search space*. Using a probabilistic interpretation where the items are assigned with a probability of relevance, a tag $t$ reduces the search space if the probability mass function $p(i|t,s)$ over the set of items $\Gamma_s$ is more concentrated than the original $p(i|s)$ (see discussion below).

*Probability estimation.* It is worth highlighting that there are many ways to estimate the probabilities of relevance $p(i|s)$ and $p(i|t,s)$. More importantly, it is *not* our goal to advocate a particular estimator, or to compare the efficacy of alternative estimators. In particular, the evaluation of our proposed method considers two possible estimators: *i)* a language model [33]; and, *ii)* a topic model based on Latent Dirichlet Allocation (LDA) [12].

### 4.2. An Information-theoretical Approach

We split the presentation of our method into three parts: first, we present how we estimate the reduction of search space by a tag; second, we discuss an approach to estimate the relevance of the set of items retrieved by tag; and, finally, we combine these two components.

*Estimating search space reduction*. In our model of exploratory search a tag reduces the search space by *leading to a higher concentration* on the probability of relevance over the set of retrieved items.

More formally, given the distribution of probability of relevance $p(\cdot|s)$, and the conditional probability distribution $p(\cdot|t,s)$ over the set of relevant items $\Gamma_s$, to measure how much information one gains by using the channel defined by a tag $t$ to read the set of items $\Gamma_s$, the proposed method uses the Kullback-Leibler divergence [8] of the two distributions:

$$D_{KL}\big(p(\cdot|t,s)\|p(\cdot|s)\big) = \sum_{i\in\Gamma_s} p(i|t,s)\log\frac{p(i|t,s)}{p(i|s)} \quad (1)$$

where $p(i|t,s)$ represents the probability that an item $i$ is relevant to a given information seeker $s$ when she uses a tag $t$ to navigate the system; while $p(i|s)$ represents the probability that an item $i$ is relevant to $s$.

Equation 1 measures the reduction in the item search space by a given tag $t$, as it quantifies how much the distribution of relevance conditional on a tag $p(i|t,s)$ diverges from the probability of relevance of an item $i$. The reduction in search space occurs, for example, when conditioning $p(i|s)$ to a tag $t$ concentrates the probability of relevance over a smaller set of items. However, as conditioning to a tag may increase the concentration of the probability mass $p(\cdot|s)$ over fewer relevant items. Therefore, it is necessary to complement Equation 1 with a measure of relevance of items a tag $t$ delivers to an information seeker $s$.

*Estimating delivered relevance*. To estimate the relevance of a set of items retrieved by tag $t$ to a particular user $s$, we compare the set of items retrieved (ordered by probability of relevance) to a reference point -- a subset with top items of $\Gamma_s$ ordered by probability of relevance. The intuition is that the more items from the top of the ranked $\Gamma_s$ the tag retrieves, the more valuable it will be. Note that according to this definition a tag maximizes its ability to retrieve relevant items by retrieving all items. This, however, does not necessarily maximize its value, as it will depend on the reduction of the search space produced by the tag (Eq. 1).

More formally, let $I^t$ be the set of items retrieved by a tag $t$ and *not* already annotated by the information seeker $s$ (i.e., $I^t \not\subset I_s$). Also, let $I^t$ be ordered by relevance to an information seeker $s$. Let $\Gamma_s^{[k]}$ be the set of top-$k$ most relevant items to $s$ from $\Gamma_s$ when ordered according to $p(\cdot|s)$. We define the relevance delivered by a tag $t$ to an information seeker $s$ as:

$$\rho(t,s) = 1 - \tau\big(I^t, \Gamma_s^{[k]}\big) \quad (2)$$

where $\tau\big(I^t, \Gamma_s^{[k]}\big)$ is the *generalized* Kendall's $\tau$ distance [9] between $I^t$ and $\Gamma_s^{[k]}$, and $k = |I^t|$. Kendall's distance measures the fraction of the number of changes needed (in regards to the maximum number of changes) to convert one rank ($I^t$) to the other ($\Gamma_s^{[k]}$). A distance of 0 means that both ranks are the same, while 1 states that the ranks are exact opposites. A penalty of 1 is incurred when items appear on one rank but not the other. The rationale is that the more relevant items a given tag retrieves, the smaller is the distance and the closer to 1 the value of $\rho(t,s)$ gets.

*Combining relevance and reduction of search space*. The final step is to define the estimate of the value of tag $t$, from the perspective of an information seeker $s$, $v(t,s)$.

DEFINITION 2. *Given an information seeker $s$ and her set of relevant items $\Gamma_s$, the value of a tag $t$ to $s$, is defined as:*

$$v(t,s) = \rho(t,s) D_{KL}\big(p(\cdot|t,s)\|p(\cdot|s)\big) \quad (3)$$

The rationale behind this definition of tag value is that if a tag $t$ retrieves only items with low relevance to $s$, the factor $\rho(t,s)$ penalizes the value, as it computes the distance from the retrieved set of items to the set of estimated relevant items to the user. Therefore, tag $t$ has little value to the information seeker, even though it may reduce the search space towards a subset of $\Gamma_s$. On the other hand, if $t$ leads the user to a subset of relevant items, its value is proportional to the reduction in search space, as the relevance of the retrieved items -- $\rho(t,s)$ -- will be close to one and will have a smaller penalty effect.

### 4.3. Properties of the Proposed Method

This section shows that the method we propose can indeed distinguish between two arbitrary tags, when they deliver different levels of relevance and reduction of search space.

*Search space reduction*. As described in the previous section, we use a probabilistic interpretation of the search space: a tag reduces the search space if the probability mass function $p(i|t,s)$ over the set of items $\Gamma_s$ is more concentrated than $p(i|s)$.

The goal of this analysis is to show that our proposed method is able to distinguish between two tags that lead to different levels of search space reduction. More formally, we prove the following proposition:

PROPOSITION 1. *Given an information seeker $s$, if a tag $t$ reduces the search space more than another tag $w$ by moving the probability mass towards more relevant items*, then $D_{KL}(p(\cdot|t,s)\|p(\cdot|s)) > D_{KL}(p(\cdot|w,s)\|p(\cdot|s))$.

PROOF. See Appendix A.

*Relevance level*. We show that, from the perspective of a given information seeker $s$, Equation 2 distinguishes two tags if they deliver two different levels of relevance. To show that our proposed method has this property, we prove the following proposition.

PROPOSITION 2. *Given an information seeker $s$, if a tag $t$ retrieves more relevant items than a tag $w$, it follows that $\rho(t,s) > \rho(w,s)$*. PROOF. See Appendix B.

### 5. VALIDATION

The previous section presents proofs that the proposed method can differentiate between two tags when they lead to different levels of search space reduction and relevance of retrieved items. This section complements these results by performing an experiment with real data to test the *accuracy* of our method. The method is accurate if the tag values it produces match users' perception of value.

Two hard constraints limit the validation experiments we can execute: we do not have access to browsing traces and we do not have access to a ground truth, that is, direct estimates of users' perception of value.

We have, however, access to tag assignment traces in a number of systems and here we use them to estimate our method's accuracy based the following intuition: when a user assigns a tag to an item, this tag had a high value for the user from the perspective of a future search for that particular object. Thus, if our method consistently estimates the value of the previously used tags higher than the value of random tags (that the user has *not* used before), then there is a strong indication that the method is accurate in quantifying tag value as perceived by users.

#### 5.1. Experiment Design

To test the hypothesis that the proposed method passes this accuracy criterion, we collect tag assignments from *LibraryThing* (http://www.macle.nl/tud/LT). Our data set consisted of 37,232 items, 10,559 tags used by 7,279 users.

The experiment consists of two major parts: *i*) finding the best probability estimator parameters (steps 1 to 4 below), which are then used as inputs to our method; and, *ii*) for each user, computing the value of tags from two sets; a sample of tags from the user's vocabulary and a sample of tags *not* in the user's vocabulary (steps 5 and 6). These sets are denoted, respectively, by $G_s \subset T_s$ and $R_s \subset T - T_s$. It is important to highlight that neither tags in $G_s$ nor tags in $R_s$ used in the parameter estimation phase. Thus, the method has no information whether the user has annotated items with specific tags before.

More formally, this experiment tests the hypothesis that the method is able to assign higher value to tags in $G_s$ (user vocabulary) than to tags in $R_s$ (random tags). Our experiment has the following steps:

1. We select a sample of users that use the system more than occasionally, that is, users with at least 50 annotated items. We denote this sample by $S_{50}$.

2. With the tagging trace sorted by annotation timestamp, we break the set of annotations $A$ into three sets: $A_{train}$, $A_{param}$, and $A_{test}$. The training set contains the first 80% (sorted by date) of items annotated for the users in the sample $S_{50}$. The validation and test set are each composed by 10% of the remaining annotations. We made sure that all tags/items on the validation and test sets, also appeared on training set.

3. We train the estimators (based on different parameters) for the probability distributions $p(i|t,s)$ and $p(i|s)$ on $A_{train}$. Models trained were based on topic models presented in [12]. As in [12], we were unable to reproduce the results in [33], thus our choice to use topic models only.

4. The set of items in $A_{param}$ are then used to measure average Success@10 of the estimator for each user. Success@10 captures the fraction of times at least one relevant item, that is, one item in the validation set, appeared in the first set of the first 10 items when sorted by $p(i|t,s)$ or $p(i|s)$. Each probability distribution is evaluated independently of the other. This way, we pick the best estimator *parameterization* for each probability distributions. The best estimators reached Success@10 values of 0.05 and 0.06 for $p(i|t,s)$ and $p(i|s)$ respectively. Parameters used are $\alpha = 0.1/|I|, \beta = 0.1/|T|, \gamma = 0.001$.

5. With the best parameterization thus obtained, we use $A_{test}$ to perform our experiments. Recall that no parameter tuning is done on this test set. Now, for each user $s \in S_{50}$, two sets of tags are constructed, namely: *hidden* and *random*. The *hidden* set, denoted by $G_s \subset T_s$, contains tags used by user $s$ in the test set $A_{test}$. The *random* set, $R_s$, is comprised of 50 tags that are randomly selected from the trace and have not been used by the user on any of: $A_{train}$ or $A_{param}$ sets.

6. Finally, we compare the distributions of tag value $v(t,s)$ for tags in $G = \bigcup_{s \in S} G_s$ to that of the tags in $R = \bigcup_{s \in S} R_s$.

### 5.2. Results

We start by showing, in Figure 2, the results of a baseline method - here referred to as *naïve method* - which simply uses the number of items the tag retrieves and the average relevance of retrieved items to compute the tag value. The plot shows the cumulative distribution functions (CDF) of values for both tag sets from the perspective of all users in the *LibraryThing* data. The result shows that the naïve method is not efficient in distinguishing between tags that users find valuable (i.e., those part of the hidden set) and the others (i.e., those part of the random set).

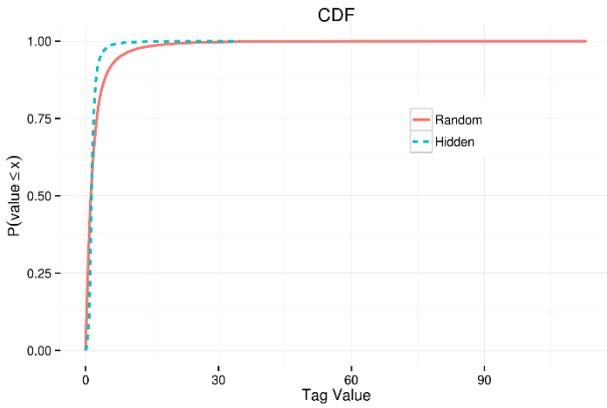

**Figure 2. Comparison between the cumulative distribution functions (CDF) of tag values (naïve method) for tags in each set (Hidden and Random), from the perspective of each user in the LibraryThing data set. D- = 0.25 ($p < 2.2 \times 10^{-16}$).**

In contrast, Figure 3 shows the CDF for tags values computed using our proposed method based on the information theoretical approach. The result shows that the distribution of tag values for tags in the *random* set $R$.

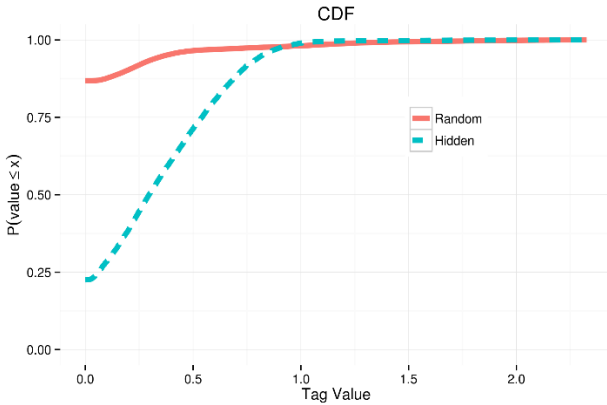

**Figure 3. Comparison between the cumulative distribution functions (CDF) of tag values (our proposed method) for tags in each set (Hidden and Random), from the perspective of each user in the LibraryThing data set. D- = 0.64 ($p < 2.2 \times 10^{-16}$).**

To confirm that the tag values for tags in one sample are significantly larger than those from the other sample, we apply a Kolmogorov-Smirnov test. In fact, the test allows the rejection of the null hypothesis that the values in the samples come from the same distribution, and accept the alternative hypothesis that the distribution of tag values for tags in the *hidden* set lies below that of *random*.

In particular, we observe that the *D*-statistic, which measures the distance between the two CDFs, for the information theoretical method is 2.5 times larger than that of the naïve method. The larger the difference the better is the method in distinguishing the valuable (hidden) from random tags. In fact, D- = 0.25 ($p < 2.2 \times 10^{-16}$) for Naïve; D- = 0.64 ($p < 2.2 \times 10^{-16}$) for our method.

Therefore, these experiments provide evidence that the proposed method (formalized by Equation 3) is accurate, as it is able to assign higher values to those tags that users perceive as more valuable.

We complement the analysis by also comparing the methods according to their achieved Mean Reciprocal Rank (MRR). The intuition is if a method rank tags from the hidden set higher than those in random, that method is more accurate. More formally, for each user $s \in S_{50}$, consider a set of tags, $T_s = G_s \cup R_s$, where tags are ranked by their value, as estimated by a given method. Also, let $r_s(t)$ be the rank of the first tag in $T_s$, such that $t \in G_s$ (i.e., it is part of the hidden set). Thus, we calculate the MRR for each method as follows: $MRR = \frac{1}{|S_{50}|} \sum_{s \in S} \frac{1}{r_s(t)}$.

By computing the MRR for each method, we observe that $MRR_{naive} = 0.04$, while $MRR_{info} = 0.19$ (i.e., the MRR achieved by our method is 4 times higher). This result suggests that our method is able to rank tags, which users have expressed interest, higher than those users have not expressed interest, although the information that tags have been used by the user is hidden from the method.

### 6. SUMMARY AND DISCUSSION

This study focuses on the problem of quantifying the value of peer-produced tags for exploratory search. To address this problem, we use a combination of qualitative and quantitative research methods. First, using semi-structured contextual interviews to collect the data and grounded theory for the analysis, this study characterizes the aspects that influence users' perception of tags' value. Second, we design a method that quantifies tag value by considering the two most salient aspects among those identified by the qualitative analysis. Finally, we perform an evaluation with real tagging data, and provide evidence that the proposed method is able to distinguish more valuable tags.

**Threats to validity**. This study is subject to some design decisions that may impact its validity: *(i) External validity:*

although the qualitative and quantitative studies are performed at small scale, which limits our ability to make generic claims about the findings, users reported their experience with a diverse set of systems; moreover, the qualitative analysis covers usage scenarios of both tag production and tag-based search in a variety of systems. We believe that this broad set of real experiences reduces the threat to external validity; *(ii) Internal validity* – one potential source of threats to internal validity is the interaction between data used in the probability estimators and the methods that assess the value of a tag. However we guarantee that this threat is removed by breaking the trace into three disjoint segments (training, parameter estimation, and test) to avoid using the same data in training (i.e., probability estimators) and testing (i.e., tag value computation).

**Future work**. Finally, it is important to note that our qualitative analysis uncovers several aspects that influence the users' perception of tag value in exploratory search. Our proposed method quantifies the two most salient ones. We plan to extend this method to account for the other aspects. A larger evaluation using either a collected ground truth or click traces is a natural extension of this study.

**APPENDIX A. PROOF OF PROPOSITION 1**

PROOF. **1st Condition**. Given an information seeker $s$, if a tag $t$ reduces the search space more than another tag $w$, we have that: $p(i|t,s)$ is more concentrated than $p(i|w,s)$, where $i \in \Gamma_s$. Therefore, $H(p(i|t,s)) < H(p(i|w,s))$, where $H$ is Shannon's entropy [8].

**2nd Condition**. Moreover, if a tag $t$ moves the probability mass towards more relevant items than the tag $w$ does, this means that there are at least two items in $j, k \in \Gamma_s$ where $p(j|s) > p(k|s)$ such that when conditioning the probability to tag $t$ and $w$, respectively, we have that $p(j|t,s) > p(j|w,s)$ and $p(k|t,s) < p(k|w,s)$. Note that, to conserve the probability mass, it necessary that $|p(j|t,s) - p(j|w,s)| = |p(k|t,s) - p(k|w,s)|$.

Putting these two conditions together and applying Equation 1 to $p(\cdot|t,s)$ and $p(\cdot|w,s)$, we prove, by contradiction, that the proposition holds:

$$D_{KL}(p(i|t,s)\|p(i|s)) < D_{KL}(p(i|w,s)\|p(i|s))$$

$$\sum_{i \in \Gamma_s} p(i|t,s) \log \frac{p(i|t,s)}{p(i|s)} < \sum_{i \in \Gamma_s} p(i|w,s) \log \frac{p(i|w,s)}{p(i|s)}$$

$$\sum_{i \in \Gamma_s} p(i|t,s) \log p(i|t,s) - \sum_{i \in \Gamma_s} p(i|t,s) \log p(i|s)$$

$$< \sum_{i \in \Gamma_s} p(i|w,s) \log p(i|w,s)$$

$$- \sum_{i \in \Gamma_s} p(i|w,s) \log p(i|s)$$

Replacing the first summations by the entropy leads to:

$$-H(p(i|t,s)) - \sum_{i \in \Gamma_s} p(i|t,s) \log p(i|s)$$

$$< -H(p(i|w,s)) - \sum_{i \in \Gamma_s} p(i|w,s) \log p(i|s)$$

Next, we expand the second summation with:

$$-H(p(i|t,s)) - \sum_{i \in \Gamma_s - \{j,k\}} p(i|t,s) \log p(i|s) - p(j|t,s) \log p(j|s)$$

$$- p(k|t,s) \log p(k|s)$$

$$< -H(p(i|w,s)) - \sum_{i \in \Gamma_s - \{j,k\}} p(i|w,s) \log p(i|s)$$

$$- p(j|w,s) \log p(j|s) - p(k|w,s) \log p(k|s)$$

Cancelling the equal summations from both sides, leads to:

$$-H(p(i|t,s)) + \log p(j|s) < -H(p(i|w,s)) + \log p(k|s)$$

$$H(p(i|w,s)) - H(p(i|t,s)) < \log p(k|s) - \log p(j|s)$$

From the first condition set forth in the proposition, we know that $H(p(i|w,s)) - H(p(i|t,s)) > 0$, and from the second condition $p(k|s) - p(j|s) < 0$. Therefore, the last equation contradicts the original conditions, and the propositions holds.

**APPENDIX B. PROOF OF PROPOSITION 2**

PROOF. If $t$ retrieves more relevant items than $w$, we have that:

$$\tau\left(I^t, \Gamma_s^{[k]}\right) < \tau\left(I^w, \Gamma_s^{[k']}\right)$$

where $k = |I^t|$ and $k' = |I^w|$. By inverting the signs and adding 1 to both sides, we have:

$$1 - \tau\left(I^t, \Gamma_s^{[k]}\right) > 1 - \tau\left(I^w, \Gamma_s^{[k']}\right)$$

Therefore, $\rho(t,s) > \rho(w,s)$.

**APPENDIX C. CONTEXTUAL INTERVIEW GUIDELINES**

The contextual interview guide consisted of the questions below. Note that although the interviews help to collect data that enable us to confirm previous studies about both motivations to use tags and the types of use, the primary goal of this investigation is to understand what aspects influence the users' perception of value when choosing tags during information seeking tasks:

1. Why do you use tags? Why do you use each of these specific systems you mentioned?
2. What's the perceived value of tags produced by other users to you?
3. Can you describe search interfaces/systems of your choice that you use when looking for a set of items related to the same topic? For example, to explore a given topic of interest. (*Probes*: to find articles related to a topic of interest)
4. What are the situations where you feel the search interfaces mentioned above are more adequate to perform your search tasks, as opposed to other alternatives? (*Probes*: traditional keyword-based search vs. AND-search navigation)?
5. Please, describe/show us (in as much details as possible) the process you follow when using exploratory search. You can recount your last experience, for example.
6. Consider a scenario where you are looking for content on a given topic of interest. How do you choose among tags when navigating (i.e., performing information seeking tasks)?
7. Can you show us an example of an exploratory search where you had to choose among tags to proceed?
8. Why did you choose these tags while looking up these content items (from question 7)?
9. How does the partial search results influence the tags you choose to proceed with the navigation?
10. Let's talk about a different use of tags: annotation instead its use in search/navigation. How did you choose the tags when annotating content?
11. Do you speak/write/read more than one language? If so, how do these multiple languages influence your choice of tags?
12. How the intended use of content you found during your search/navigation influence your choice of tags to annotate it?

In the second part, users are requested to 'solve' the following navigation tasks:

- **Task 0** (tutorial). Find articles related to *cooking*. (The goal is to get the user acquainted to the *Getboo* interface and enable her to perform task 1 and 2 without much intervention).
- **Task 1**. Find articles related to your *work* that are interesting to you (and new).
- **Task 2**. Find articles related to your *hobbies* that are interesting to you.

We note that the search tasks are deliberately vague. The reason is that such tasks are the ones that motivate users to go into exploratory search mode [27] rather than trying to locate a single specific answer to an information need (e.g., what is a factotum? What is the *blindekuh* restaurant's location in Zürich?)

**APPENDIX D. SOURCE CODE**

Source code is available at: http://hidden, and http://hidden. Both repositories contain information on how to reproduce our results and further discussions on implementations choices we made.